\documentstyle[12pt,epsf]{article}
\renewcommand{\bar}[1]{\overline{#1}}

\newcommand{\mathbf}{{\bf}}
\newcommand{\mathrm}{{\rm}}

\begin{document}

\bigskip\bigskip
{\centerline{\large
\bf 
ON THE ULTRAVIOLET DIVERGENCE IN QED
}}

\vspace{22pt}

\centerline{
\bf Ji Sun}

\vspace{8pt}

{\centerline {Department of Technical Physics, Peking University,
Beijing, P.R.~China}


\vspace{10pt}
\begin{center} {\large \bf Abstract}

\end{center}

The well-known physical equivalence drawn from hole theory
is applied in this article.  The author suggests to replace,
in the part of Feynman diagram which cannot be fixed by
experiments, each fermion field operator, and hence fermion
propagator, by pairs of equivalent fermion field operators
and propagators.  The formulation of this article
thus yields additional terms which reveal characteristic
effects that have not been explored previously; such
characteristic effects lead to the appearence of logarithmic
running terms and that finite radiative corrections are
directly obtained in calculations.

\vfill
\centerline{
}

\vfill

\centerline{Published in Hadronic Journal 21 (1998) 583-612.}
\newpage

\vfill
\vspace{40pt}

\begin{center} {\large \bf About the Author}
\vspace{15pt}

\end{center}

	{\bf Ji Sun} (1921-1997), one of the influential physicists in
China, has
been dedicated in research and education of physics for half of a
century. He graduated from the Department of Physics, Shanghai Jiaotong
University, Shanghai, China in 1947. Since then, he had engaged in
research and education in quantum machenics and particle physics in high educational
institutions of China like Nankai University, Tsinghua University and
Peking University. He was one of the founder and leading researcher of the
Department of Technical Physics in Peking University. In the last ten
years of his life, he had struggled with prostate cancer. The tremendous pain
and even the high level paralysis caused by the disease did not prevent
him from pursuring scientific truth until the last minute of his life. His
dedication to physics and scientific world would be remembered by
generations to come.

} 

\vfill

\newpage

\section{Introduction}

Ultraviolet divergence is an important problem in QED.  
Investigation on this problem, to expose more characteristics 
of the divergence, may be helpful in the developments of
the quantum field theory \cite{pap1}.  This article is an attempt
on this problem.  In this article, we start from the well-known
and well-established physical equivalence drawn from Dirac's
hole theory \cite{pap2}, which gives also pairs of physically equivalent
fermion propagators.  In general, as measurements or observations
fix only one of a physically equivalent pair, the physical
equivalence from the hole theory thus gives nothing new,
so such discussion is superfluous.  However, there are
certain parts of Feynman diagrams in QED processes in which
the fermion fields or fermion propagators cannot be fixed
experimentally ( e.g. fermion propagators in a self energy loop).
Both of the two equivalent propagators are equally probable
to happen; thus the physical equivalence might lead to
additional term or terms.  It will be shown in section 3
that the new additional terms coming from physical equivalence
really reveal new characteristic physical effects which are
closely related with ultraviolet divergences and can yield
finite radiative corrections and hence finite results
in direct calculations of QED processes.

It might be quite interesting to note that one of
the striking characteristics of the ultraviolet divergence,
i.e. the appearence of, for example, logarithmic running of
QED coupling constant with scale, which has been verified
in precise electroweak measurements, is also given by the
formulation of this article, as will be shown in Sec.~3;
the logarithmic running terms really appear in this article;
however, there are, meanwhile, really more such logarithmic
running terms with different charges.
Such logarithmic terms will combine to give finite radiative corrections.

As the first paper of this work, main focus is given to the fundamental
assumption and formulation, together with their foundations.
Here as illustration, one QED process, the vertex, is calculated;
other QED processes will be given in subsequent papers.

\section{Fundamental Assumption and 
the Foundation of the Formulation}

\subsection{Preliminary discussions on hole theory}

In order to propose the fundamental assumption,
we need first to investigate the hole theory \cite{pap2}.
Before the investigation on hole theory, we review first
the characterisitics of tensors formed by fermion
and boson field operators under the transformation
$x \to -x$, given by refs.~\cite{pap4} and \cite{pap5}; their relevant
results are rearranged\footnote{
(\ref{eq1a}),(\ref{eq1b}) can be obtained from relevant 
results in refs.~\cite{pap4}, \cite{pap5}:
Consider for example the case of boson fields $U$ which can
be divided into two classes $U^+$ and $U^-$; From the paper of Pauli,
the fundamental homogeneous linear equation of the fields,
which should be in the typical form by the requirement of
invariance against proper Lorentz group, as
\renewcommand{\theequation}{\arabic{equation}P}
\begin{equation}
\sum k U^+=\sum U^-, ~~~ \sum k U^-=\sum U^+
\label{eq1P}
\end{equation}
There are altogether two possible substitutions keeping 
Eq.~(\ref{eq1P})
invariant:
\begin{equation} 
k_l \to - k_l (k_l=-i\frac{\partial}{\partial x_l}, l=1,2,3,4.), 
U^+ \to U^+, U^- \to - U^-.            
\label{eq2P}
\end{equation}
(as $U^*$ belong to same class as $U$; for simplicity, we use $U$ to
represent both $U$ and $U^*$ here and below. )  and
\renewcommand{\theequation}{\arabic{equation}'P}
\setcounter{equation}{1}
\begin{equation}
k_i \to -k_i, U^+ \to - U^+, U^- \to U^-
\label{eq2P'}
\end{equation}
As the propagation vector $k_i$ belongs to $U^-$ class, 
so only (\ref{eq2P})
among the two substitutions (\ref{eq2P}), (\ref{eq2P'}) is consistent;
which may also be considered as: when  $k_i \to - k_i$, 
if Eq.~(\ref{eq1P}) is required
to remain invariant,  then $U^+ \to U^+$, $U^- \to -U^-$; and hence
$T \to T$, $S \to -S$.  As the invariance
of field equation (\ref{eq1P}) is a fundamental requirement,
so we may always write:  
If $x_{\mu} \to -x_{\mu}$, then $T \to T$, $S \to -S$. 
which is (\ref{eq1b}).  Similarly for (\ref{eq1a}).
} 
in the form:

Under  the transformation $x_{\mu} \to -x_{\mu}$\footnote{Fermion fields 
here mean free fermion fields; non-free fields
may include a boson term, e.g. $p + e/c\, A$.  The fermion field $p$
and the boson field term $A$ obeys 
(\ref{eq1a}) and (\ref{eq1b}) respectively.}:
\renewcommand{\theequation}{\arabic{equation}a}
\setcounter{equation}{0}
\begin{equation}  
        {\mathrm fermions~ fields:} ~ T \to  -T, ~~ S \to S.            
\label{eq1a} 
\end{equation}
\renewcommand{\theequation}{\arabic{equation}b}
\setcounter{equation}{0}
\begin{equation}                        
        {\mathrm bosons~ fields:}  ~  T  \to T,  ~~ S  \to -S.                         
\label{eq1b}
\end{equation}
$T$, $S$ in (\ref{eq1a}), ( in (\ref{eq1b})) are tensors formed  by fermion
(boson) field operators.  $T$ represents tensors of even rank,
including scalars, skew symmetric and symmetric tensors
of second rank such as energy momentum tensor, etc.  $S$ represents
tensors of odd rank,  including vectors such as the charge current
density vector, etc.

Now, as a preliminary to the fundamental assumption, we discuss
the physical equivalences drawn from hole theory.

We begin our discussion by {\it reexpressing  the well known
equivalences drawn from  the hole theory}:
\renewcommand{\theequation}{\arabic{equation}}
\setcounter{equation}{1}
\begin{eqnarray}
&                b_{-\vec{p},r=3,4} \equiv  d^{\dag}_{\vec{p},r=1,2};  
&                b^{\dag}_{-\vec{p},r=3,4} \equiv d_{\vec{p},r=1,2};  
\nonumber \\
&                b_{\vec{p},r=1,2} \equiv d^{\dag}_{-\vec{p},r=3,4}; 
&                b^{\dag}_{\vec{p},r=1,2} \equiv d_{-\vec{p},r=3,4}
\label{eq2}                 
\end{eqnarray}
The equivalences may also be written as  
$$\sum_{\vec{p},r=3,4} b_{-\vec{p},r} u^{(r)}(-\vec{p})=
\sum_{\vec{p},r=1,2} d^{\dag}_{p,r} v^{(r)}(\vec{p})$$ etc.
Here $b$ ($d$) refer to $-e$ ($+e$) fermions, $\equiv$ 
denotes physical equivalence,
the both sides of which express the same physical entity or process.

Now we combine the equivalences (\ref{eq2}) with (\ref{eq1a}).  
As from (\ref{eq1a}) that
for fermions, $p(\vec{p},E)$ reverses its sign on reversing 
$x(\vec{x},t)$,
it is reasonable to associate $p(\vec{p},E)$ with $x(\vec{x},t)$,
(and hence $-p(-\vec{p},-E)$ with $-x (-\vec{x},-t)$) 
to agree with experimental
facts.  We obtain immediately a time dependent expression of
the equivalences drawn from hole theory.  
The example $b_{-\vec{p},r=3,4} \equiv d^{\dag}_{\vec{p},r=1,2}$
is now expressed as the equivalence between the processes (a) (b)
in Fig.~1; namely, there are two equivalent processes:  
(a) an electron
with charge $-e$, momentum $-\vec{p}$, energy $-E$ 
propagating in $-t$ sense,
denoted here and here after as $(-e,-\vec{p},-E,-t)$, is annihilated
at $-x(-\vec{x},-t)$. ( $-x(-\vec{x},-t)$ is 
just the space-time point $x(\vec{x},t)$
viewed in the frame $-x$.)  (b) an electron with 
$(+e, \vec{p}, E, t)$
is created at $x(\vec{x},t)$.  (a) and (b) are two expressions of
a same physical process.  All other equivalences in (\ref{eq2}) can
be reexpressed similarly.
%

\begin{figure}[htbp]
\begin{center}
\leavevmode {\epsfysize=12.0cm \epsffile{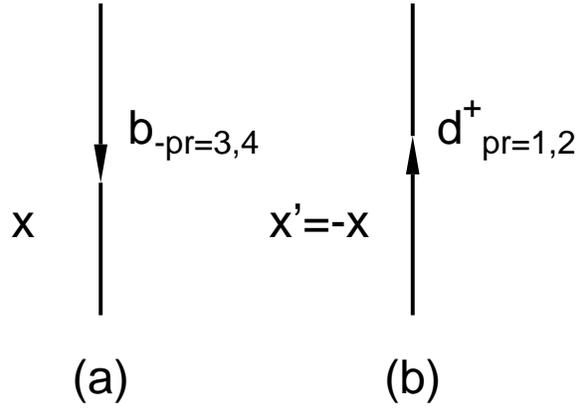}} 
\end{center}
\caption[*]{\baselineskip 20pt 
Equivalent pair of fermion field operators: 
(a),  annihilation of $(-e,-\vec{p},-E)$;  (b), creation of 
$(+e,\vec{p},E)$. 
$x$ 
($x'$)  being  reference frame of coordinate of $-e$  ($+e$)  electron; 
($x'=-x$).                        
}
\label{sunf1}
\end{figure}

The equivalences drawn from hole theory  also lead  directly
to pairs of equivalent electron propagators as depicted
in Fig.~2, provided the equivalences  hold at both  $x_1$ and $x_2$.
Thus there are  two physically equivalent propagators in Fig.~2:
Fig.~2 (a) is an electron with $-e,\vec{p},E$, propagating from $x_1$ to 
$x_2$ in $+t$ sense, denoted as $(-e,\vec{p},E,t)$; while Fig.~2 (b)
is an electron with $+e,-\vec{p},-E$  propagating from $x'_2=-x_2$
to $x'_1=-x_1$, in $-t$ sense., denoted as $(+e,-\vec{p},-E,-t)$.
Mathematical forms of Fig.~2 (a), (b) will
be given below.
%

\begin{figure}[htbp]
\begin{center}
\leavevmode {\epsfysize=12.0cm \epsffile{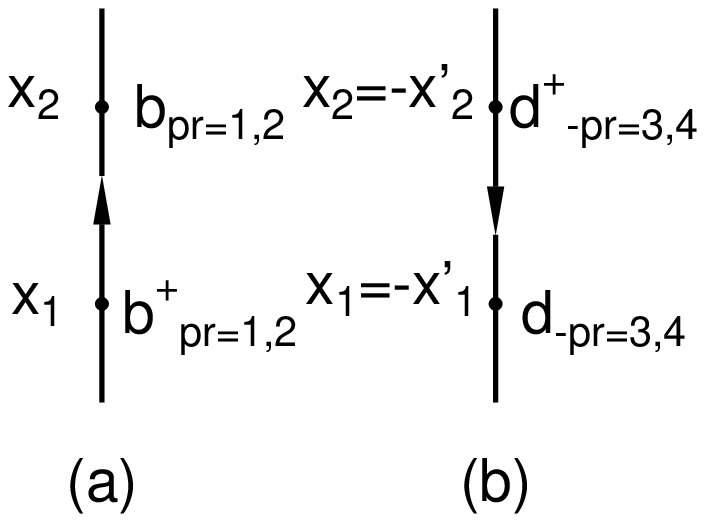}} 
\end{center}
\caption[*]{\baselineskip 20pt 
Equivalent pair of fermion propagators: (a),  
$(-e,\vec{p},E,\vec{x},t)$;    
(b),   $(+e,-\vec{p},-E,-\vec{x},-t)$. $x$  ($x'$)  being  reference 
frame of coordinate of $-e$ ($+e$) electron;  ($x'=-x$).          
}
\label{sunf2}
\end{figure}

There are thus equivalences between $(-e,\vec{p},E,t)$ and 
$(+e,-\vec{p},-E,-t)$
processes.  Relations between the equivalent fermion pairs (a),
(b). both of Fig.~1 and of Fig.~2 are: 
(a)$\to$(b) and (b)$\to$(a) will occur under
the simultaneous reflections ($x_{\mu} \to -x_{\mu}$) 
and ($Q \to -Q$).  Note that $p_{\mu} \to -p_{\mu}$ 
is just
a consequence of $x_{\mu} \to -x_{\mu}$ by 
(\ref{eq1a}); and ($Q \to -Q$) is a consequence of transpose,
which interchanges the initial and final states (see below, 2.2.1) and 2)).

The pair (a),(b) of Fig.~2, for example, may be considered as one
fermion propagator, which is (a) or $(\vec{p}, E;  \vec{x}, t)$, 
if it is regarded
as $-e$ propagator; while it is (b) or $(-\vec{p}, -E;  -\vec{x},
-t)$, if it is
regarded as $+e$ propagator.  Thus it follows that the axes of
reference {\it frames} of {\it physical equivalent} 
$-e$ and $+e$ fermion fields
(particle and antiparticle) are opposite in sense.

\subsection{Physically equivalent pairs of fermion field operators}

\noindent
1) The definition

The well known equivalences drawn from Dirac's hole theory
can be formulated or generalized as:  " {\it To each  fermion field
operator $\Psi(x)$  there is a $\Psi^R(x)$, 
which is physically equivalent to $\Psi(x)$,
defined as
\begin{equation}
\Psi^R(x)=R \Psi(x) R^{-1}=\Psi'^{T}(x)
\label{eq3}
\end{equation}
where $R=r\,tr$,  $r$ being the reflection ($x \to -x$) operator, 
which bring $\Psi(x)$  
into $\Psi'(x)=O \Psi(-x)$, $O$ being a matrix, $tr$ 
being the transpose operator,
which bring $\Psi(x)$ into $\Psi^T(x)$, 
with the initial and final states interchanged.
}"  

\noindent
2) Several notes:

\noindent
i)  The operator $O$ given here is a matrix, which is proved, under
the requirement of invariance of Dirac equation, etc, as 
$O=\xi \gamma_5$.
($\xi= \pm1, \pm i$) \cite{pap5}.  Thus,
\renewcommand{\theequation}{\arabic{equation}a}
\setcounter{equation}{2}
\begin{equation}
\Psi^R(x)=\xi \gamma_5 \Psi^T(-x)
\label{eq3a}
\end{equation}
which shows that the transformation of $R$ is just the same as that of
joint operation $CPT$.

\noindent
ii) Eq.~(\ref{eq3a}) shows that the charge of
$\Psi^R(x)$  should be $-Q$ if that of $\Psi(x)$ is $Q$,
since they are connected by the transformation $CPT$.  
Eq.~(\ref{eq3}) gives
directly that $\Psi^R(x)$ is the fermion proceeding 
in the sense of time $-t$,
if $\Psi(x)$ is that proceeding in the sense of time $+t$; 
the opposite sense
of time gives the interchange of creation and annihilation operators.
Thus, e.g., the creation of $-e$ charge of $\Psi(x)$  
is turned into annihilation
of $-e$, or creation of $+e$ charge of $\Psi^R(x)$; 
therefore $\Psi^R(x)$  and $\Psi(x)$  are opposite
in charge.  This is just the physical equivalence drawn from hole theory.

The transformation (\ref{eq3}) or (\ref{eq3a}) is 
called "{\bf strong reflection}"
by Pauli \cite{pap3} (see also \cite{pap5}).

\noindent
iii) Thus, if we take $\Psi(x)$  as a $Q=-e$ fermion with $+p(+E)$ 
proceeding
in the sense of $+t$, i.e., $\Psi_{-e,p,+E,+t}$, 
$\Psi^R(x)$ will then be  $\Psi^R_{+e,-p,-E,-t}$ (see also
Eq.~(\ref{eq1a})).
The physical equivalence between $\Psi_{-e,p,+E,+t}$  
and $\Psi^R_{+e,-p,-E,-t}$  is just the equivalences
in Eq.~(\ref{eq2}) (e.g. 
$b^{\dag}_{\vec{p},r=1,2} \equiv d_{-\vec{p},r=3,4}$ 
etc., $b$, $d$ being operators of $-e$, $+e$ respectively,
remember that the transpose makes $\Psi^R$ to 
proceed in the sense $-t$.).

\noindent
iv) The fact that $-E, -t$ fermions are closely related to
the $+E,+t$ fermions with opposite charge was already given
by Feynman \cite{pap6}.  However, the problem is treated here in a
somewhat different point of view; in this article, 
{\it all processes
of fermion field operators and of products of them always proceed
in the order of increasing time in its "own frame of reference"}
(namely the frame that the $t$-axis is directed in $+t$ ($-t$) sense for
the fermion $\Psi_{+t}$ ($\Psi{-t}$) ); $-t$ 
{\it arises only when a process is viewed in other
frame of reference}.  This explains also why "a particle travelling
from $x_1$ to $x_2$ is the same as an 
antiparticle travelling from $x_2$ to $x_1$."
(see also Fig.2).

The transpose operation in (\ref{eq3}) and (\ref{eq3a}) 
keeps that all processes
proceed in the order of increasing of time.

\noindent
3) It is significant to note that it has been proved that the
transformation of $R$ (Eq.~(\ref{eq3})) {\it leaves both field equations and
commutation relations invariant \cite{pap7}}.  
Also it has been proved that,
if the electromagnetic fields $A_{\mu}$ is transformed by $R$ 
simultaneously,
i.e.,
\renewcommand{\theequation}{\arabic{equation}'}
\setcounter{equation}{2}
\begin{equation}
A_{\mu}^{R}(x) = R A_{\mu}(x) R^{-1}=A'^{T}_{\mu}(x)
\label{eq3'}
\end{equation}
the transformation of $R$ {\it leaves both field equations and commutation
relations of fermion and boson fields invariant}.  Moreover,
{\it the fundamental equations of quantum electrodynamics are
invariant}
under the transformation of $R$ \cite{pap7}.

Therefore the transformation of $R$ is consistently 
defined for full interacting quantum field theory.

\subsection{The Fundamental Assumption}

\noindent
1)  Before  proposing  the fundamental assumption,  it  might  be 
significant to notice that QED processes consist of two  possible 
cases  (or  parts of Feynman diagram): (a). observable  case  (or 
parts),   which  can  be  directly  observed  or  determined   by 
measurements,  (e.g.  external  fermion or  boson  lines).   (b). 
unobservable case (or parts), which cannot be directly determined 
by measurements or observations (e.g. fermion propagators in  the 
loops of self energy processes).

        In  case  (a), the observable case,  although  there  are 
equivalent  expressions $\Psi$ and $\Psi^R$, 
a fermion  field  or  a  physical 
process is always uniquely expressed by only one expression fixed 
by  measurement  (e.g. $\Psi$, if it is expressed 
with $+E, +t$ which 
agrees  with  measurement condition).  Everything  is  as  usual, 
nothing  new can be given by the physical equivalence defined  in 
Eq.~(\ref{eq3}).

        In  case  (b),  the unobservable case,  a  fermion  field 
operator or physical process cannot be determined by measurements 
or  observations.  {\it No measurement can determine which  expression 
should be taken}.  The {\it two equivalent expressions}, 
e.g.  $\Psi$ and $\Psi^R$,  are 
then {\it equally probable}.

        It  should  be noted that: the case (a) is  the  ordinary 
case, in which each $C$, $P$, $T$ reflection can be operated  separately; 
while  in case (b), although the transformation $R$  satisfies  $CPT$ 
theorem, single reflections $P$, $T$, $C$ cannot be performed  separately 
in general, since no observation or measurement can be  performed 
in  case  (b);  hence a single reflection,  e.g., $T$,  cannot  be 
defined.  Only case (a) has been considered nowadays; while  case 
(b) has never been investigated so far (case (b) appears only  in 
some intermediate steps.).

        In  case  (b), the physically  equivalent  fermion  field  
$\Psi(x)$ and $\Psi^R(x)$,  which  satify  
the same quantum field  equation  with  e.m. 
interaction   and  commutation  relation,  are   observed   under 
different conditions (e.g. $\Psi(x)$ is measured or viewed in the frame of 
reference $+x (+\vec{x}, +t)$, $Q= -e$; 
while $\Psi^R(x)$ is viewed in the frame 
$-x (-\vec{x},  -t)$, $Q=+e$).  
In each measurement, we can {\it only} measure  either 
$\Psi(x)$ or $\Psi^R(x)$, {\it but not both}; 
so it is suitable to apply,  as  in  quantum 
mechanics,  the {\it superposition principle}.  
As no  measurement  can 
determine which of $\Psi$ and $\Psi^R$ 
is more probable than the other; so $\Psi$ and $\Psi^R$ should  appear 
with   equal  probability;  we  arrive  thus  naturally  at   the 
{\it fundamental assumption}. 

\noindent
2) Fundamental assumption:

        In  the  unobservable case (case (b)),  a  fermion  field 
operator,  which  is  $\Psi(x)$  in the observable  case  (case  (a)),  is 
expressed as
\renewcommand{\theequation}{\arabic{equation}}
\setcounter{equation}{3}
\begin{equation}
\frac{1}{\sqrt{2}} \left[ \Psi(x)+\Psi^R(x) \right]
\label{eq4}                                                                
\end{equation}
{\it where $\Psi^R(x)=R \Psi(x) R^{-1}$ is defined in Eq.~(\ref{eq3}) 
and  the  statements  in  that 
paragraph.  This is the fundamental assumption of this article.}

        It   should  be  noted  that,  in  the  case  (a), 
$\Psi^R(x)$    is 
automatically  turned  into $\Psi(x)$ 
which is expressed in the  frame  of 
reference  $+x  (+\vec{x}, +t)$.  
This is due to that $\Psi(x)$, $\Psi^R(x)$   are  just  two 
expressions  of  the same $-e$ fermion field viewed from  frame  of 
reference $+x (+\vec{x}, +t)$, $Q=-e$ and $-x (-\vec{x}, -t)$, 
$Q=+e$,  respectively; 
in other words,  $\Psi^R(x)$ is just  $\Psi(x)$ viewed 
in the frame $-x (-\vec{x}, -t)$,  $Q=+e$.  
The  case (a) is, e.g., to observe $\Psi^R(x)$, 
which should be  transformed 
to the frame $+x (\vec{x}, t)$, i.e. into $\Psi(x)$, as
\begin{equation}
R \Psi^R(x) R^{-1} \left | O \right>=\Psi(x) \left|O \right>           
\label{eq5}
\end{equation} 
$\left | O \right>$ is the observable state 
which is fixed in the frame $+x (+\vec{x}, +t)$.

The  transformation  (\ref{eq5}) between two expressions  of  the 
same  field leaves the fermion field  unchanged.   Transformation 
(\ref{eq5}) is just twice operation of transformation $R$ on 
$\Psi(x)$.

As $\Psi(x)$ and $\Psi^R(x)$ in expression (\ref{eq4}) 
should be  measured  in  two 
different measurements, if we pass from case (b) to case (a), the 
norm  becomes,  by  (\ref{eq5})  and (\ref{eq4}), 
$\frac{1}{2}\left[ |\Psi(x)|^2+|\Psi^R(x)|^2 \right]=
|\Psi(x)|^2$.   Thus,  each  fermion  field 
operator  is well connected between unobservable  and  observable 
parts  of a Feynman diagram; this {\it guarantees the conservation  of 
fermion number} and everything of fermion fields.

        In  the  case  (b), as $\Psi(x)$ and $\Psi^R(x)$ 
are  equally  probable,  the 
complete  set  of fermion field operators should be  extended  to 
that given by (A1) and (A1') in Appendix A.

\subsection{The $S$-matrix formulation}

        The  formulation  of this article is the same as  the  
$S$-matrix formulation of conventional QED, i.e.
\begin{eqnarray}        
&S=&\sum_n S^{(n)} 
\nonumber \\
&S^{(n)}=&\frac{(-i)^n}{n!}\int_{-\infty}^{\infty}{\mathrm d}^4 x_{n}
\int_{-\infty}^{\infty}{\mathrm d}^4 x_{n-1} 
\cdots
\int_{-\infty}^{\infty}{\mathrm d}^4 x_{1} 
\nonumber \\
&& T \left[ {\mathbf H}_I(x_n) {\mathbf H}_I(x_{n-1})\cdots
{\mathbf H}_I(x_1) \right]          
\label{eq6}
\end{eqnarray}
{\it except  only that in the unobservable case (case (b)),}   
$\Psi(x)$ {\it in  each} 
${\mathbf H}_I(x)=-i e\bar{\Psi}(x)\gamma_{\mu}\Psi(x)
A_{\mu}(x)$ {\it is replaced by} $\frac{1}{\sqrt{2}}
\left[\Psi(x)+\Psi^R(x) \right]$ {\it given in (\ref{eq4}). 
$A_{\mu}(x)$ multiplied with $\Psi^R(x)$ is correspondingly
transformed by $R$, ((\ref{eq3'})), since $\Psi$ interacts with
$A$ at same space-time points
(while in case (a) nothing  is 
changed}).

        In (\ref{eq6}),  ${\mathbf H}_I(x)$ 
denote interaction Hamiltonian  densities.  
The  chronological  operator $T$ is defined as usual.  (  i.e.  all 
factors  in the bracket after $T$ are arranged so that the time  of 
the  factors  are  increasing from right to left; a  factor
$\delta_p$   is 
multiplied where $p$ is the number of permutations of fermion field 
operators to bring them into chronological order).

        In case (b), due to the replacement of 
$\Psi(x)$ by $\frac{1}{\sqrt{2}}
\left[\Psi(x)+\Psi^R(x) \right]$, each ${\mathbf H}_I(x)$ 
in Eq.~(\ref{eq6}) is then replaced by,  
$\frac{1}{2}\left[{\mathbf H}_I(x) +
{\mathbf H}_I^R(x) \right]$, 
${\mathbf H}_I^R(x) = -i e\bar{\Psi}^R(x)\gamma_{\mu}\Psi^R(x)
A_{\mu}^R(x)$.

        Actually  ${\mathbf H}_I(x)$ becomes  
$\frac{i e}{2} \left[ \bar{\Psi}(x)\gamma_{\mu}\Psi(x) A_{\mu}(x)+
\bar{\Psi}^R(x)\gamma_{\mu}\Psi^R(x) A_{\mu}\right]$
on replacing $\Psi(x)$ by $\frac{1}{\sqrt{2}}\left[\Psi(x)+\Psi^R(x) \right]$; 
all  other 
terms that might appear in the direct replacement are excluded by 
the   requirement  that  every  term  in  ${\mathbf H}_I(x)$  
should   be   an 
electromagnetic  interaction term which is a product  of  various 
fields  at  the same space-time point.  Therefore, in  case  (b), 
${\mathbf H}_I(x)$ is replaced by  $\frac{1}{2}\left[{\mathbf H}_I(x) +
{\mathbf H}_I^R(x) \right]$. 
Also, if we pass from case (b) to  case 
(a),  as  each $\Psi^R(x)$ is automatically turned into $\Psi(x)$,   
$\frac{1}{2}\left[{\mathbf H}_I(x) +
{\mathbf H}_I^R(x) \right]$ is  turned  into  
${\mathbf H}_I(x)$ automatically.

The product of  $\frac{1}{2}\left[{\mathbf H}_I(x) +
{\mathbf H}_I^R(x) \right]$ is arranged by the chronological operator 
$T$  into  two products ( the product of ${\mathbf H}_I(x)$ and
${\mathbf H}^R_I(x)$).   It  may  be 
interesting  to note that the order of factors in the product  of 
${\mathbf H}_I(x)$ is automatically reversed.

\subsection{Summarization of the formulation of this article}

        I).   The  formulation  of  this  article  differs   from 
conventional  QED only in one point, namely: 
{\it In the  unobservable 
case, (or parts of a Feynman diagram; the case (b)), each fermion 
field operator $\Psi(x)$ is replaced by 
$\frac{1}{\sqrt{2}}\left[\Psi(x)+\Psi^R(x) \right]$, 
(the fundamental assumption).

        Everything except this point is the same as  conventional 
QED and hence need not be qualified or discussed.}

        II).  The critical point of the replacement stated in  I) 
is the transformation of $R$ drawn from hole theory ((\ref{eq3}) and 
(\ref{eq3'})).

        1). It has been proved that the fundamental equations  of 
quantum electrodynamics are invariant under the transformation  $R$ 
\cite{pap7}.  {\it Therefore the transformation of  $R$ of  this article  leaves 
the  fundamental  equations  of  QED  invariant,  and  hence   is 
consistently  defined for full interacting quantum field theory.}

        2).  {\it Causality is satisfied}, which can be seen  directly 
{\it by  the  chronological operator $T$ in (\ref{eq6})}.   
Also,  {\it all  processes}  
defined above {\it proceed in the sense of increasing time};  $-t$ arises 
{\it only  when}   a  process {\it is viewed in other  frame}  of  reference.  
Furthermore,  if only observable states defined above, such as
$\Psi(x)$, 
are considered, there appears only $+t$.

        3) {\it Unitarity is satisfied.}
 
                   i)  The  transformation  $R$  defined  in   this 
article  is  itself  unitary.   Actually,  the  transformation  $R$ 
satisfies $CPT$ theorem.  Therefore the formulation of this article 
satisfies unitarity.

           ii) The fermion(s) (and fermion propagator(s)) in each 
$S^{(n)}$ in (\ref{eq6}) {\it are the same as in}
$S^{(n)}$ of the corresponding conventional $S$-matrix theory,
throughout each process except that $\Psi(x)$  is replaced by  
$\frac{1}{\sqrt{2}}\left[\Psi(x)+\Psi^R(x) \right]$ 
in the observable case (case(b)).

        a) In the expression 
$\frac{1}{\sqrt{2}}\left[\Psi(x)+\Psi^R(x) \right]$ 
the sum of probabilities of $\Psi(x)$ and $\Psi^R(x)$ , 
which cannot be measured simultaneously, is equal to unity.

        b) The two expressions $\Psi(x)$ and 
$\Psi^R(x)$ of the same fermion field  in 
Eq.~(\ref{eq4}) can only be measured in two different measurements; 
so if 
we pass from case (b) to case (a), $\Psi^R(x)$ should be 
transformed into $\Psi(x)$, 
as given by Eq.~(\ref{eq5}).  
So the norm of $\frac{1}{\sqrt{2}}\left[\Psi(x)+\Psi^R(x) \right]$ 
is then $\frac{1}{2}\left[|\Psi(x)|^2+|\Psi^R(x)|^2
\right]=|\Psi(x)|^2$ 
which is just that 
of $\Psi(x)$ in the case (a).  Therefore, each fermion field operator  is 
well  connected  between unobservable and observable parts  of  a 
Feynman diagram.  This guarrantees the unitarity and conservation 
of quantum number and everything of fermion fields.

        4)   The   transformation  $R$  is  also   called   "strong 
reflection"  by  Pauli \cite{pap3} (see also \cite{pap5}),  
which  satisfies  $CPT$ 
theorem.   However,  it  should be noted  that  separate  $P$, $T$,... 
reflections, which is defined only in observable case (case (a)), 
is not defined in the unopbservable case (case (b)).

\section{Consequences}

The  fundamental assumption and the formulation given  in 
Sec.~2 lead directly to the following consequences:

\noindent
1. {\it Two equivalent pairs of fermion propagators.}

        We  have already given in Sec.~2 that the occurrence  of 
two  physically  equivalent  expressions of  each  fermion  field 
operator leads to the occurrence of:

        (1)   two   equivalent  expressions   of   each   fermion 
propagator.   This  is  a  direct  consequence  of  the  physical 
equivalence  drawn from hole theory, as shown in Fig.~2.   There 
are  in  general  two equivalent  propagators  for  each  fermion 
propagator in Feynman diagram.  In the case when a propagator can 
be  determined by measurement (the observable case (a)), the  two 
equivalent propagators will automatically become identical.

        (2) two equivalent pairs of fermion propagators.

        If  we include the unobservable case (b), the  appearence 
of pairs of physically equivalent fermion field operators extends 
the complete set of fermion field operators to (A1), (A1')  given 
in  Appendix  A.   Such extended complete set  of  fermion  field 
operators  can  constitute  four fermion  propagators,  i.e. {\it two 
equivalent   pairs   of   fermion   propagators}.    Namely,   the 
propagator(A)  formed by (A1a),(A1'a) and its equivalent  (B)  by 
(A1b),(A1'b);  and similarly (C) and (D) formed  respectively  by 
(A1c),(A1'c) and (A1d),(A1'd),  as given in Appendix A.  
The results of Appendix  A,  the 
propagators (A),(B),(C),(D), are written here:                                                          
\renewcommand{\theequation}{\Alph{equation}}
\setcounter{equation}{0}
\begin{equation}
\frac{i}{(2\pi)^4}\int {\mathrm d}^4 p \frac{-i \hat{p}+m}{p^2+m^2}
e^{i p(x_2-x_1)}
\label{eqA}
\end{equation}
\begin{equation}
\frac{i}{(2\pi)^4}\int {\mathrm d}^4 p \frac{i \hat{p}-m}{p^2+m^2}
e^{i p(x_2-x_1)}
\label{eqB}
\end{equation}
\begin{equation}
\frac{i}{(2\pi)^4}\int {\mathrm d}^4 p \frac{-i \hat{p}-m}{p^2+m^2}
e^{i p(x_2-x_1)}
\label{eqC}
\end{equation}
\begin{equation}
\frac{i}{(2\pi)^4}\int {\mathrm d}^4 p \frac{i \hat{p}+m}{p^2+m^2}
e^{i p(x_2-x_1)}
\label{eqD}
\end{equation}

\noindent
2.  Occurrence of characteristic effects in the  calculations  of 
typical divergence problems in QED.

        From  the statement in 1, in the unobservable  case  (b), 
the  formulation  Eq.~(\ref{eq6})  gives  directly,  for  each   fermion 
propagator in the Feynman diagram of a QED process:

        (1)  a sum of two propagators of an equivalent pair  each 
multiplied with its interacting e.m. field operator.

        (2) two equivalent pairs of fermion propagators (\ref{eqA}), 
(\ref{eqB}) 
and (\ref{eqC}), (\ref{eqD}), each pair occurs with equal probability.

        (1) and  (2)  give two significant characteristic effects 
which will be discussed in and after the illustrative examples of 
QED process given below.   In order to illustrate the fundamental 
assumption and the formulation of the article, typical divergence 
processes have been calculated.  As the calculations are somewhat 
lengthy, we choose only one of them, the vertex, here to show the 
ability of the formulation of this article; other processes  will 
be given in subsequent papers.

\noindent
{\bf The vertex} \cite{pap8}

        The  $S$-matrix  element  of  the  vertex  of  an  electron 
incoming at $x_1$ with $(-e,\vec{p}_1,E_1)$    
outgoing at  $x_2$ with $(-e,\vec{p}_2,E_2)$    
and interacting with the 
external field $A_{\mu}^e(x_3)=A^e_{\mu} e^{i q x_3}$ 
at $x_3$, is $\left<f \right|S^{(3)}\left|i\right>$.  
The $S^{(3)}$  in Eq.~(\ref{eq6}) has been  rewritten, 
in Appendix B.,1., as
\renewcommand{\theequation}{\arabic{equation}}
\setcounter{equation}{6}
\begin{eqnarray}
&&S^{(3)}=-\frac{e^3}{8}\int_{-\infty}^{\infty}{\mathrm d}^4 x_2
\int_{-\infty}^{\infty}{\mathrm d}^4 x_3
\int_{-\infty}^{\infty}{\mathrm d}^4 x_1
T \{
\nonumber \\
&&
\big(\bar{\Psi}(x_2)\hat{A}^{\cdot}(x_2)
\Psi^{{\mathbf \cdot}}(x_2)\bar{\Psi}^{{\mathbf \cdot}}(x_3)
+\bar{\Psi}^R(x_2)\hat{A}^{R \cdot \cdot}(x_2)
\Psi^{R {\mathbf \cdot \cdot}}(x_2)\bar{\Psi}^{R {\mathbf \cdot \cdot}}(x_3)
\big)
\hat{A}_{\mu}^{e}(x_3)
\nonumber \\
&&
\big(\Psi^{{\mathbf \cdot \cdot \cdot}}
(x'_3)\bar{\Psi}^{{\mathbf \cdot \cdot \cdot}}(x'_1)
\hat{A}^{\cdot}(x_1)
\Psi(x'_1)
+
\Psi^{R {\mathbf \cdot \cdot \cdot \cdot}}
(x'_3)\bar{\Psi}^{R {\mathbf \cdot \cdot \cdot \cdot}}(x'_1)
\hat{A}^{R \cdot \cdot}(x_1)
\Psi^R(x'_1)
\big)
\}
\label{eq7}
\end{eqnarray}
                                                                                                                                               
        The terms which are zero in the matrix element
$\left<f\right|S^{(3)}\left|i\right>$ have  not 
been written, where
$\left|i\right>=b^{+}_{\vec{p}_1,r_1}\left|\,\right>_0$, 
$\left|f\right>=b^{+}_{\vec{p}_2,r_2}\left|\,\right>_0$;  
the factor $\frac{1}{3!}$  is  omitted since 
only  one  of the $3!$ possible figures is  taken.  The  space-time 
variables  in (\ref{eq7}) are only $x_2$, $x_3$, $x_1$; 
$x'_3$, $x'_1$ in (\ref{eq7}) are just  $\pm x_3$,  $\pm x_1$,  
the  sign  is $+$ or $-$ according to the fermion  propagators  taken 
(see Appendix A.2).

        The  external field $\hat{A}^e_{\mu}(x)$ in (\ref{eq7}) 
is equally well be  multiplied 
to the second square bracket; namely written as  
$\hat{A}^e_{\mu}(x'_3)$ instead of $\hat{A}^e_{\mu}(x_3)$.
 
        As  pairs  of  fermion  propagators  in  the  two  square 
brackets in (\ref{eq7}) should be taken over all possible pairs,  (A),(B) 
and (C),(D),  there are  thus four cases:

\noindent
I) .  (A),(B) in both square brackets.

\noindent
II).  (A),(B) in the first, (C),(D) in the second square bracket .

\noindent
III). (C),(D) in the first,(A),(B) in the second square bracket.

\noindent
IV).  (C),(D) in both square brackets.

        As it can be easily shown that (III), (IV) give the  same 
value  as  (I), (II); so it is only necessary to  calculate  (I), 
(II) with the result times 2.  So Eq.~(\ref{eq7}) can be written as 
\begin{equation}
        S^{(3)}=-\frac{e^3}{4}
\int_{-\infty}^{\infty}{\mathrm d}^4 x_2 
\int_{-\infty}^{\infty}{\mathrm d}^4 x_3 
\int_{-\infty}^{\infty}{\mathrm d}^4 x_1 
T(\{I\}+\{II\})
\end{equation}
Where 
\renewcommand{\theequation}{\arabic{equation}'1a}
\setcounter{equation}{7}
\begin{eqnarray}
&{I} =&\{\left[ \big(\bar{\Psi}(x_2)\hat{A}^{\cdot}(x_2)(A)_{x_2,x_3}
+\bar{\Psi}^R(x_2)\hat{A}^{R \cdot
\cdot}(x_2)(B)_{x_2,x_3}\big)\right] \hat{A}^e_{\mu}(x_3)
\nonumber \\
&&\left[\big( 
(A)_{x'_3,x'_1}\hat{A}^{\cdot}(x'_1)\Psi(x'_1)               
+(B)_{x'_3,x'_1}\hat{A}^{R \cdot \cdot}(x'_1)\Psi^R(x'_1)
\big)\right]
\}       
\label{eq8'1a}
\end{eqnarray}
\renewcommand{\theequation}{\arabic{equation}'2a}
\setcounter{equation}{7}
\begin{eqnarray}
&{II}=&\{
\left[ \big(\bar{\Psi}(x_2)\hat{A}^{\cdot}(x_2)(A)_{x_2,x_3}
+\bar{\Psi}^R(x_2)\hat{A}^{R \cdot
\cdot}(x_2)(B)_{x_2,x_3}\big)\right] \hat{A}^e_{\mu}(x_3)
\nonumber \\
&&\left[\big( 
(C)_{x'_3,x'_1}\hat{A}^{\cdot}(x'_1)\Psi(x'_1)      
+(D)_{x'_3,x'_1}\hat{A}^{R \cdot \cdot}(x'_1)\Psi^R(x'_1)
\big)\right]
\}
\label{eq8'2a}
\end{eqnarray}
$(A)_{x_2,x_3}$ being  $\Psi^{\cdot}(x_2)\bar{\Psi}^{\cdot}(x_3)$
with propagator (A); similarly for propagators (B),(C),(D).

        {I} and {II} are equally written as:
\renewcommand{\theequation}{\arabic{equation}'1b}
\setcounter{equation}{7}
\begin{eqnarray}
&{I} =&\{\left[ \big(\bar{\Psi}(x_2)\hat{A}^{\cdot}(x_2)(A)_{x_2,x_3}
+\bar{\Psi}^R(x_2)\hat{A}^{R \cdot
\cdot}(x_2)(B)_{x_2,x_3}\big)\right] \bar{A}^e_{\mu}(x'_3)
\nonumber \\
&&\left[\big( 
(A)_{x'_3,x'_1}\hat{A}^{\cdot}(x'_1)\Psi(x'_1)                                       
+(B)_{x'_3,x'_1}\hat{A}^{R \cdot \cdot}(x'_1)\Psi^R(x'_1)
\big)\right]
\}       
\label{eq8'1b}
\end{eqnarray}
\renewcommand{\theequation}{\arabic{equation}'2b}
\setcounter{equation}{7}
\begin{eqnarray}
&{II}=&\{
\left[ \big(\bar{\Psi}(x_2)\hat{A}^{\cdot}(x_2)(A)_{x_2,x_3}
+\bar{\Psi}^R(x_2)\hat{A}^{R \cdot
\cdot}(x_2)(B)_{x_2,x_3}\big)\right] \bar{A}^e_{\mu}(x'_3)
\nonumber \\
&&\left[\big( 
(C)_{x'_3,x'_1}\hat{A}^{\cdot}(x'_1)\Psi(x'_1)   
+(D)_{x'_3,x'_1}\hat{A}^{R \cdot \cdot}(x'_1)\Psi^R(x'_1)
\big)\right]
\}
\label{eq8'2b}
\end{eqnarray}

        {III} and {IV} are defined similarly, hence,
\renewcommand{\theequation}{\arabic{equation}}
\setcounter{equation}{7}
\begin{eqnarray}
&\left<f\right| S^{(3)} \left|i\right>=&
\frac{-i e^3}{4} \bar{u}^{(r_2)}(\vec{p}_2)
\int\frac{{\mathrm d}^4 k}{k^2+\lambda^2}\gamma_{\nu}
\nonumber \\
&&[\big\{
\frac{-i(\hat{p}_2-\hat{k}_2)+m}{(p_2-k)^2+m^2}\gamma_{\mu}
\frac{-i(\hat{p}_1-\hat{k}_2)+m}{(p_1-k)^2+m^2}
\nonumber \\
&&+
\frac{i(\hat{p}_2+\hat{k}_2)-m}{(p_2+k)^2+m^2}\gamma_{\mu}
\frac{i(\hat{p}_1+\hat{k}_2)-m}{(p_1+k)^2+m^2}
\big\}
\delta^4(p_1-p_2+q) 
\nonumber \\
&&+\big\{
\frac{-i(\hat{p}_2-\hat{k}_2)+m}{(p_2-k)^2+m^2}\gamma_{\mu}
\frac{i(\hat{p}_1-\hat{k}_2)+m}{(p_1-k)^2+m^2}
\nonumber \\
&&+
\frac{i(\hat{p}_2+\hat{k}_2)-m}{(p_2+k)^2+m^2}\gamma_{\mu}
\frac{-i(\hat{p}_1+\hat{k}_2)-m}{(p_1+k)^2+m^2}
\big\}
\delta^4(p_1-p_2)]
\nonumber \\
&&
\gamma_{\nu} a_{\mu}(q) u^{(r_1)}(\vec{p}_1)
\label{eq8}
\end{eqnarray}
                                                         
        The  factor $\delta^4(p_1-p_2+q)$ 
 in the first term of 
(\ref{eq8}), is  different from the corresponding factor  
$\delta^4(p_1-p_2)$ in the second.  

        In  the first term of (\ref{eq8}), which  is 
the  integral of {I}, (\ref{eq8'1a}) or (\ref{eq8'1b}), 
in which the  propagators 
in  both square brackets are all (A), (B); so $x'_3=x_3$.  
Hence  (\ref{eq8'1a}) = 
(\ref{eq8'1b}) as required.

        While the second term of (\ref{eq8}),  which 
is  the  integral  of  (II),  (\ref{eq8'2a})  or  (\ref{eq8'2b}),  
in  which  the 
propagators in the first square bracket are (A), (B); while those 
in the second square bracket are (C), (D). So 
$x'_3=-x_3$ (see Appendix  A).  
However, it is required that (\ref{eq8'2a}) {\it has to be equal 
to} (\ref{eq8'2b}) for 
all  values  of  $x_3$; namely it is required 
that $\hat{A}^e_{\mu}(x_3)=\hat{A}^e_{\mu}(x'_3)$,  
i.e.,  it  is 
required  that $e^{i q x_3}=e^{-i q x_3}$; 
as $x_3$ {\it runs over all space-time points, so it  is 
required} 
$q=0$\footnote{Actually, $q=0$ is a significant physical condition; 
since  as 
boson  fields behaves differently from fermion fields  under  the 
reflection  $x \to  -x$ as given by (\ref{eq1a}), (\ref{eq1b}), 
the factor $e^{i p' x}$   of  a 
fermion propagator with $p'=p-e/c\,A$ 
is unchanged under $x \to -x$ if 
$A=0$ is assumed, since $e^{i(-b)(-x)}=e^{i p x}$  by (\ref{eq1a}).
}.       

        The calculation of (\ref{eq8}) 
is given in Appendix B.  The final 
is  
\begin{equation}
\Lambda^{(2)}_{\mu f} (p_1,p_2;q)=-\frac{\alpha}{\pi}
\big\{\left[(\frac{q^2}{3m^2}\ln \frac{m}{\lambda}-
\frac{q^2}{8 m^2}\right]\gamma_{\mu}
-\frac{i}{8 m} (\gamma_{\mu}\hat{q}-\hat{q}\gamma_{\mu})
 \big\}.     
\label{eq9}
\end{equation}
to order $q^2$, which is the same as that 
obtained in the renormalization theory.

\noindent
{\bf About the characteristic effects}

        The two terms in the square bracket of (\ref{eq8}) means that  we 
should  take the average of {I} and {II} which appear with  equal 
probability.  In order to take an insight into Eq.~(\ref{eq8}) consider 
for a moment the fictitious processes in which there is only  {I} 
or {II} alone, i.e.
$$
S^{(3)}=-\frac{e^3}{4}\int_{-\infty}^{\infty}{\mathrm d}^4 x_2
\int_{-\infty}^{\infty}{\mathrm d}^4 x_3
\int_{-\infty}^{\infty}{\mathrm d}^4 x_1
T\{I\} 
$$
$$
S^{(3)}=-\frac{e^3}{4}\int_{-\infty}^{\infty}{\mathrm d}^4 x_2
\int_{-\infty}^{\infty}{\mathrm d}^4 x_3
\int_{-\infty}^{\infty}{\mathrm d}^4 x_1
T\{II\} 
$$

        We have, from (A4), (A4I), (A4II) in Appendix B,
\begin{eqnarray}
&&\big(\left<f\right| S^{(3)} \left|i\right>\big)_I=
\frac{-i e^3}{2} \bar{u}^{(r_2)}(\vec{p}_2)
\int {\mathrm d}^4 k  \int_0^1 {\mathrm d} x \int_0^x {\mathrm d} y
\nonumber \\
&&[\frac{\{(2-2x-x^2)m^2-\frac{k^2}{2}+(1-x+y)(1-y)q^2\}\gamma_{\mu}
}
{\{k^2+m^2x^2+q^2y(x-y)+\lambda^2(1-x)\}^3}
\nonumber \\
&&+\frac{im q_{\mu}(1+x)(2 y-x) + mx (1-x) \sigma_{\mu\nu}q_{\nu}}
{\{k^2+m^2x^2+q^2y(x-y)+\lambda^2(1-x)\}^3}
]a_{\mu}(q) u^{(r_1)}(p_1)
\nonumber
\end{eqnarray}
\begin{eqnarray}
&&\big(\left<f\right| S^{(3)} \left|i\right>\big)_{II}=
\frac{-i e^3}{2} \bar{u}^{(r_2)}(\vec{p}_2)
\int {\mathrm d}^4 k  \int_0^1 {\mathrm d} x \int_0^x {\mathrm d} y
\nonumber \\
&&\frac{\{-(2-2x+x^2)m^2+\frac{k^2}{2}\}\gamma_{\mu}}
{\{k^2+m^2x^2+\lambda^2(1-x)\}^3}
a_{\mu}(q) u^{(r_1)}(p_1)
\nonumber
\end{eqnarray}
    
        We  see at once that there are logarithmic running  terms 
in both fictitious processes 
$\big(\left<f\right| S^{(3)} \left|i\right>\big)_I$ and 
$\big(\left<f\right| S^{(3)} \left|i\right>\big)_{II}$ 
(due to the $\pm \frac{k^2}{2}$ terms in the 
numerators of both integrands.).  This shows why the conventional 
formulation  of  QED, which includes  
$\big(\left<f\right| S^{(3)} \left|i\right>\big)_I$ only,  gives  logarithmic 
running  terms.   While  in  the  formulation  of  this  article,  
althrough there appear also logarithmic terms, the 
$\pm \frac{k^2}{2}$ terms in the 
numerators  of  (\ref{eqA4I}),  and (\ref{eqA4II})  
conceled  each  other  before 
integration  over  ${\mathrm d}^4 k$;  
thus  there  is  no  divergence  in  all 
processes of this article; 
the finite radiative corrections are obtained directly. 

        We see that the first pair of propagators in $\{I\}$ and
$\{II\}$ 
are  the same, they differ only in the second pair, (A),  (B)  in 
$\{I\}$,  while  (C),  (D)  in $\{II\}$.   As  (A),(B)  and  (C),(D)  are 
propagators $(-e,\vec{p},E,\vec{x},t)$, 
$(+e,-\vec{p},-E,-\vec{x},-t)$ and  
$(-e,-\vec{p},-E,-\vec{x},-t)$, 
$(+e,\vec{p},E,\vec{x},t)$ respectively, 
so, e.g., the propagators $(\vec{p},E,\vec{x},t)$ in 
(A),(B)  and  in (C),(D) are respectively (A) and (D)  which  are 
{\it opposite}  in charge.  Similarly, the two propagators  
$(-\vec{p},-E,-\vec{x},-t)$  
are also {\it opposite}  in charge.  
So the {\it charge content  of  the 
pair (A),(B) is different  from that  of (C),(D). } 

        Therefore  the  terms  $\{I\}$ and $\{II\}$  
in  (\ref{eq8})  are  really 
coexisting  physical processes with different charges.  So it  is 
not surprising that we can measure logarithmic running of , e.g., 
coupling  constant , while the two coexisting processes  
$\{I\}$  and 
$\{II\}$  have  the effect of concellation with each  other  at  very 
large  $k$  or  very  small  distance (see  
the $\pm \frac{k^2}{2}$ terms  in  the 
numerators  of (\ref{eqA4I}) and (\ref{eqA4II})) 
which renders the final  results 
finite.

        So  far we have discussed, in the process of the  vertex, 
the characteristic effect coming from the averaging of coexisting 
$\{I\}$  and $\{II\}$ processes.  
There is another characteristic  effect 
coming from replacing a fermion propagator interacting with  e.m. 
(photon)  field  by two propagators of an  equivalent  pair  each 
multiplied with its interacting e.m. field operator.  This effect 
will cancel the logarithmic divergence in the self energy of free 
electrons; while for non-free electrons, we obtain, on  combining 
this effect with the averaging of coexisting processes 
$\{I\}$ and $\{II\}$ 
discussed  above,  finite radiative corrections.  These  will  be 
given by the processes in subsequent papers.

\section{Conclusion and Discussions}

        1.  Based on the fundamental assumption, $\Psi(x)$  
is replaced by $$\frac{1}{\sqrt{2}}\left[\Psi(x)+\Psi^R(x) \right]$$ 
 in case (b), the formulation of this article leads, in the  case 
(b), to that: 

        1) every propagator in Feynman diagram is replaced by two 
propagators  of  an  equivalent pair, each  multiplied  with  its 
interacting e.m. field operator.  

        2) there occur two pairs of fermion propagators, (A), (B) 
and (C), (D), each with equal probability.

        Each  of  1), 2) gives a characteristic effect  which  is 
closely related to the ultraviolet divergence.

        2.  As  given above, we have obtained, on one  hand,  the 
appearence of logarithmic  running terms; while on the other, the 
finite  results  of  calculations.  This is due  to  that,  there 
appear, in this article, logarithmic running terms with different 
charges  ( see the statements near the end of Sec.~3), so  the 
cancellation  of logarithmic terms at very large values of k,  or 
very  small  distance, is a very natural  physical  effect.   The 
cancellation  of logarithmic terms takes place before  the  final 
integration over  $k$;  
so  no ultraviolet divergences  appear  finally  in  our 
calculations; finite radiative corrections are obtained in direct 
calculations.   It  might  be possible  that,  according  to  the 
author, the appearence of ultraviolet divergence might be due  to 
the negligence of the case (b).

        3.  Since the calculations of each process  are  somewhat 
lengthy, only one of them can be given here,  Actually the author 
have calculated the three typical divergence problems, i.e.,  the 
self  energy  of  electron, the self  energy  of  photon  (vavuum 
polarization), and the vertex.  Preliminary results show that all 
the  radiative  corrections  are the same as  those  obtained  in 
conventional  renormalization  treatments.  Such  works  will  be 
given in subsequent papers.

        The  three  typical  divergences are the  source  of  all 
ultraviolet  divergences in QED; so it might be expected that  in 
the  treatment of this article, the results of all  higher  order 
diagrams  may be all the same as those given by the  conventional 
renormalization  treatment.   However,  such  works,  which   are 
somewhat  lengthy,  can only be given in a series  of  subsequent 
papers.

\vspace{20pt}
\noindent
{\bf ACKNOWLEDGMENTS}
        The  author would like to express his profound  thank  to 
Dr.  Sining Sun  and Dr. Peixin Liu for assisting in   manuscript 
preparation in many respects.

\appendix
\section{Appendix A}

\noindent
1.The complete set of fermion field operators

        Since  the  unobservable  case (b) is  included  in  this 
article, the expansion of $\Psi$ and of $\bar{\Psi}$ 
should be written, to  include 
explicitly physically equivalent fermion field operators, as
\renewcommand{\theequation}{A\arabic{equation}}
\setcounter{equation}{0}
\begin{equation}
\Psi=\frac{1}{\sqrt{2}}(\Psi_{-e,p,x}+\Psi_{+e,-p,-x}+\Psi_{-e,-p,-x}
+\Psi_{+e,p,x})  
\label{eqA1}
\end{equation}
where,
\renewcommand{\theequation}{A\arabic{equation}a}
\setcounter{equation}{0}
\begin{equation}
\Psi_{-e,p,x}=\sum_{p,r=1,2}b_r(\vec{p})u^{(r)}(\vec{p})e^{i p x}
\label{eqA1a}
\end{equation}
\renewcommand{\theequation}{A\arabic{equation}b}
\setcounter{equation}{0}
\begin{equation}
\Psi_{+e,-p,-x}=\sum_{p,r=3,4}d_r^+(-\vec{p})v^{(r)}(-\vec{p})e^{i p x}
\label{eqA1b}
\end{equation}
\renewcommand{\theequation}{A\arabic{equation}c}
\setcounter{equation}{0}
\begin{equation}
\Psi_{-e,-p,-x}=\sum_{p,r=3,4}b_r(-\vec{p})u^{(r)}(-\vec{p})e^{i p x}
\label{eqA1c}
\end{equation}
\renewcommand{\theequation}{A\arabic{equation}d}
\setcounter{equation}{0}
\begin{equation}
\Psi_{+e,p,x}=\sum_{p,r=1,2}d_r^+(\vec{p})v^{(r)}(\vec{p})e^{i p x}
\label{eqA1d}
\end{equation}

\renewcommand{\theequation}{A\arabic{equation}'}
\setcounter{equation}{0}
\begin{equation}
\bar{\Psi}=\frac{1}{\sqrt{2}}
(\bar{\Psi}_{-e,p,x}+\bar{\Psi}_{+e,-p,-x}+\bar{\Psi}_{-e,-p,-x}
+\bar{\Psi}_{+e,p,x})  
\label{eqA1'}
\end{equation}
where,
\renewcommand{\theequation}{A\arabic{equation}'a}
\setcounter{equation}{0}
\begin{equation}
\bar{\Psi}_{-e,p,x}=\sum_{p,r=1,2}b^+_r(\vec{p})
\bar{u}^{(r)}(\vec{p})e^{-i p x}
\label{eqA1'a}
\end{equation}
\renewcommand{\theequation}{A\arabic{equation}'b}
\setcounter{equation}{0}
\begin{equation}
\bar{\Psi}_{+e,-p,-x}=\sum_{p,r=3,4}d_r(-\vec{p})\bar{v}^{(r)}(-\vec{p})
e^{-i p x}
\label{eqA1'b}
\end{equation}
\renewcommand{\theequation}{A\arabic{equation}'c}
\setcounter{equation}{0}
\begin{equation}
\bar{\Psi}_{-e,-p,-x}=\sum_{p,r=3,4}b_r^+(-\vec{p})\bar{u}^{(r)}(-\vec{p})
e^{-i p x}
\label{eqA1'c}
\end{equation}
\renewcommand{\theequation}{A\arabic{equation}'d}
\setcounter{equation}{0}
\begin{equation}
\bar{\Psi}_{+e,p,x}=\sum_{p,r=1,2}d_r(\vec{p})\bar{v}^{(r)}(\vec{p})
e^{-i p x}
\label{eqA1'd}
\end{equation}

\noindent
2. Fermion propagators. 

        There  can  be formed, from the complete set  of  fermion 
field operators, (A1) and (A1'), two equivalent pairs of  fermion 
propagators, i.e. the propagators formed by (A1a), (A1'a) and  by 
(A1c),(A1'c) together with their respective equivalents formed by 
(A1b),   (A1'b)  and  by  (A1d),  (A1'd).   Now  we  derive   the 
mathematical forms of the four fermion propagators:      

        We  derive  first the mathematical form  of  the  fermion 
propagator  formed  by  (A1a),(A1'a)  and  its  equivalent  (  by 
(A1b),(A1'b)),  i.e. Fig.~2 (a) and (b). The propagator  (a),  the 
electron propagator with $(-e,\vec{p},E,t)$ has already been given in the 
usual way as the matrix element of  
$T\left(\Psi(x_2)\bar{\Psi}(x_1)\right)$
between vacuum states, where 
$\Psi(x)=\sum_{p,r=1,2}b_{p,r}u^{(r)}(\vec{p})\exp (i px)$,  
($x_{02} > x_{01}$), as
\renewcommand{\theequation}{A\arabic{equation}a}
\setcounter{equation}{1}
\begin{equation}
_0\left<\right|
T\left(\Psi(x_2)\bar{\Psi}(x_1)\right)
\left|\right>_0=
\frac{1}{(2\pi)^3}\int {\mathrm d}^3 p \frac{-i\hat{p}+m}{2E}
e^{i p(x_2-x_1)}
\label{eqA2a}
\end{equation}

        (\ref{eqA2a}) is the propagator $(-e,\vec{p},E,\vec{x},t)$ 
in its own frame  of 
reference,  called  the frame of $-e$;  it is also  the  propagator 
$(+e,\vec{p},E,\vec{x},t)$  in the frame of $+e$, 
since (\ref{eqA2a}) is  independent  of 
sign  of $Q$. To obtain the propagator (b) of Fig.~2, we may  write 
first the propagator $(+e,-\vec{p},-E,-\vec{x},-t)$  
in the frame of $+e$ by operating $r$ ($x_{\mu} \to - x_{\mu}$  
on (\ref{eqA2a}), as
\renewcommand{\theequation}{A\arabic{equation}b'}
\setcounter{equation}{1}
\begin{eqnarray}
&&_0\left<\right|
T\left(\Psi^r(x_2)\bar{\Psi}^r(x_1)\right)
\left|\right>_0
=-
_0\left<\right|
T\left(\bar{\Psi}^r(x_1)\Psi^r(x_2)\right)
\left|\right>_0
\nonumber \\
&&=\frac{1}{(2\pi)^3}\int {\mathrm d}^3 p \frac{-i\hat{p}-m}{2E}
e^{i p(x_2-x_1)}
\label{eqA2b'}
\end{eqnarray}

Mathematical forms of  an 
equivalent  pair  of propagators ( i.e.  an  equivalent  pair  of 
expressions of one propagator) should be written in same frame of 
reference since they have to be calculated together.  So we  have 
to  write  the propagator (b) as  $(+e,-\vec{p},-E,-\vec{x},-t)$  
in  the  same 
frame as (a), i.e. in the frame of $-e$.  As the axes of frames  of 
reference  of physically equivalent $-e$ and $+e$ fermion fields  are 
opposite in sense, as stated at the end of Sec.~2.1, we have only 
to replace  $x$ in (\ref{eqA2b'}) by $-x$, 
and hence $p$ by $-p$ according to  (\ref{eq1a}), 
which  leads to the required propagator $(+e,-p,-E,-x,-t)$  
in  the 
frame of $-e$, as
\renewcommand{\theequation}{A\arabic{equation}b}
\setcounter{equation}{1}
\begin{equation}
_0\left<\right|
T\left(\Psi^R(x_2)\bar{\Psi}^R(x_1)\right)
\left|\right>_0=
-\frac{1}{(2\pi)^3}\int {\mathrm d}^3 p \frac{-i\hat{p}+m}{2E}
e^{i p(x_2-x_1)}
\label{eqA2b}
\end{equation}

        Note  that  for  a  fermion  field  operator,  e.g.,  the 
equivalence  field  $\Psi^R(x)$ of $\Psi(x)$  
is just $\Psi(x)$ when viewed from $-e$  frame  $(+x,+t)$;  
however,  the equivalent propagator (\ref{eqA2b})  is  not  exactly 
(\ref{eqA2a}) when viewed in the frame of $-e$; 
but differ in a minus sign.  
This  is due to that a propagator consists of two operators,  one 
production  and  one  annihilation operator.   If  the  frame  of 
reference is changed, the order of the two operators should  also 
be  changed, and hence a minus sign is brought in.  (Notice  that 
there are now two
 operations,  the  transpose  operation  and  the  operation   of 
chronological   operator   $T$.)   The   pair   of   equivalent 
propagators  (\ref{eqA2a}),  (\ref{eqA2b}) should, 
as in usual field  theory,  be 
written in four dimensional form, denoted as (A),(B), 
\renewcommand{\theequation}{\Alph{equation}}
\setcounter{equation}{0}
\begin{equation}
\frac{i}{(2\pi)^4}\int {\mathrm d}^4 p \frac{-i \hat{p}+m}{p^2+m^2}
e^{i p(x_2-x_1)}
\end{equation}
\begin{equation}
\frac{i}{(2\pi)^4}\int {\mathrm d}^4 p \frac{i \hat{p}-m}{p^2+m^2}
e^{i p(x_2-x_1)}
\end{equation}

        We proceed next to propagator formed by (A1c),(A1'c)  and 
its  equivalent  (  by (A1d),(A1'd)).  This  equivalent  pair  of 
fermion  propagators,  called  (C),(D),   can  be  more   quickly 
obtained from (A),(B) through $x_{\mu} \to -x_{\mu}$  as:     
\begin{equation}
\frac{i}{(2\pi)^4}\int {\mathrm d}^4 p \frac{-i \hat{p}-m}{p^2+m^2}
e^{i p(x_2-x_1)}
\end{equation}
\begin{equation}
\frac{i}{(2\pi)^4}\int {\mathrm d}^4 p \frac{i \hat{p}+m}{p^2+m^2}
e^{i p(x_2-x_1)}
\end{equation}

         Each of (A),(B),(C),(D) includes, in the usual way,  two 
three dimensional propagators; e.g. (A) includes: $(-e,\vec{p},E)$ 
for $x_{02} > x_{01}$, 
and  $(+e,\vec{p},E)$ for $x_{01} > x_{02}$.  
Similarly for (B),(C),(D).   The  space-time 
variable  of propagator (A2a) is chosen as $x$, which is also  that 
of  (A). In this way the space-time variables of  (A),(B),(C),(D) 
are $x$, $-x$, $-x$, $x$  respectively . 

                (C),(D)  should  be  considered  as   independent 
propagators, although they could be obtained from (A),(B) through 
($x_{\mu} \to -x_{\mu}$),  
since we are not permitted to make the transformation 
($x_{\mu} \to -x_{\mu}$),  e.g., 
on (A) alone, which is only a part of Feynman diagram of a  whole 
QED  process.   There  are altogether  two  equivalent  pairs  of 
propagators   (A),(B)   and  (C),(D).    The   four   propagators 
(A),(B),(C),(D) given here are all written in the frame of $-e$.

\section{Appendix B}

\noindent
1.  The step from $S^{(3)}$ in (\ref{eq6}) to (\ref{eq7}).

        We write first, by (\ref{eq6})
\renewcommand{\theequation}{A\arabic{equation}}
\setcounter{equation}{2}
\begin{eqnarray}
S^{(3)}&=&-\frac{e^3}{2\sqrt{2}}
\int_{-\infty}^{\infty}{\mathrm d}^4 x_2
\int_{-\infty}^{\infty}{\mathrm d}^4 x_3
\int_{-\infty}^{\infty}{\mathrm d}^4 x_1
T\big\{ 
\nonumber \\
&&\left[\bar{\Psi}(x_2)\gamma_{\mu}
\Psi^{{\mathbf \cdot}}(x_2)
A^{\cdot}_{\mu}(x_2)
+\bar{\Psi}^R(x_2)\gamma_{\mu}
\Psi^{R {\mathbf \cdot \cdot}}(x_2)
A^{R \cdot \cdot}_{\mu}(x_2)\right]
\nonumber \\
&&\left[\bar{\Psi}^{{\mathbf \cdot}}(x_3)\gamma_{\mu}
\Psi^{{\mathbf \cdot \cdot \cdot}}(x_3)
A^{e}_{\mu}(x_3)
+\bar{\Psi}^{R {\mathbf \cdot \cdot}}(x_3)\gamma_{\mu}
\Psi^{R {\mathbf \cdot \cdot \cdot \cdot}}(x_3)
A^{e R}_{\mu}(x_3)\right]
\nonumber \\
&&\left[\bar{\Psi}^{{\mathbf \cdot \cdot \cdot}}(x_1)\gamma_{\mu}
\Psi(x_1)
A^{\cdot}_{\mu}(x_1)
+\bar{\Psi}^{R {\mathbf \cdot \cdot \cdot \cdot}}(x_1)\gamma_{\mu}
\Psi^{R }(x_1)
A^{R \cdot \cdot}_{\mu}(x_1)\right]
\big\}
\label{eqA3}
\end{eqnarray}
                                           
        The external field $A^e_{\mu}(x)$ 
is a classical e.m. field, which  can 
be   observed  or  measured  directly;  there  are  no   physical 
equivalence, no creation or annihilation of photons; 
so $A^{e R}_{\mu}(x)$ is  just 
$A^{e}_{\mu}(x)$,  and hence can 
be extracted from the square bracket.   (\ref{eqA3})  is 
then  written as (\ref{eq7}).  
(If $CPT$ operations are applied, it can  be 
seen that the twice minus sign cancelled out, 
leaving $A^{e R}_{\mu}(x)=A^{e}_{\mu}(x)$.)

        We have written the space-time variable as  in the  first 
square  bracket in (\ref{eq8}), while as $x$ in the second. 
This is  due  to 
that the pair of propagator in each square bracket may be  either 
taken  as (A),(B) or as (C),(D).  For example, if (A),(B) are  in 
both square brackets (case I)), $x'=x$;  if (A),(B) in the first  while 
(C),(D) in the second ( case II)) $x'=-x$.

\noindent
2. The steps from (\ref{eq8}) to (\ref{eq9}).

        Eq.~(\ref{eq8}) is
\renewcommand{\theequation}{\arabic{equation}}
\setcounter{equation}{7}
\begin{eqnarray}
&\left<f\right| S^{(3)} \left|i\right>=&
\frac{-i e^3}{4} \bar{u}^{(r_2)}(\vec{p}_2)
\int\frac{{\mathrm d}^4 k}{k^2+\lambda^2}\gamma_{\nu}
\nonumber \\
&&[\big\{
\frac{-i(\hat{p}_2-\hat{k}_2)+m}{(p_2-k)^2+m^2}\gamma_{\mu}
\frac{-i(\hat{p}_1-\hat{k}_2)+m}{(p_1-k)^2+m^2}
\nonumber \\
&&+
\frac{i(\hat{p}_2+\hat{k}_2)-m}{(p_2+k)^2+m^2}\gamma_{\mu}
\frac{i(\hat{p}_1+\hat{k}_2)-m}{(p_1+k)^2+m^2}
\big\}
\delta^4(p_1-p_2+q) 
\nonumber \\
&&+\big\{
\frac{-i(\hat{p}_2-\hat{k}_2)+m}{(p_2-k)^2+m^2}\gamma_{\mu}
\frac{i(\hat{p}_1-\hat{k}_2)+m}{(p_1-k)^2+m^2}
\nonumber \\
&&+
\frac{i(\hat{p}_2+\hat{k}_2)-m}{(p_2+k)^2+m^2}\gamma_{\mu}
\frac{-i(\hat{p}_1+\hat{k}_2)-m}{(p_1+k)^2+m^2}
\big\}
\delta^4(p_1-p_2)]
\nonumber \\
&&
\gamma_{\nu} a_{\mu}(q) u^{(r_1)}(\vec{p}_1)
\end{eqnarray}

        Start now from (\ref{eq8}).  
We need only to calculate the  first 
terms  in  each curly bracket in (\ref{eq8}),  
called (\ref{eq8})$_a$;   the  second 
terms  in each curly bracket can be obtained from (\ref{eq8})$_a$ 
through $k \to -k$.  
It is readily proved that the second terms  give the same results 
as the first, i.e. (\ref{eq8})$_a$;  
so  the result of (\ref{eq8}) is twice that  of 
(\ref{eq8})$_a$.
 
        As the numerator and the denominator of the first term in 
the first curly bracket
$$
\gamma_{\nu}\frac{-i(\hat{p}_2-\hat{k}_2)+m}{(p_2-k)^2+m^2}\gamma_{\mu}
\frac{-i(\hat{p}_1-\hat{k}_2)+m}{(p_1-k)^2+m^2}
\gamma_{\nu}
\delta^4(p_1-p_2+q)
$$
are respectively $\{2m^2\gamma_{\mu}-\hat{q}\gamma_{\mu}\hat{q}
+\hat{k}\gamma_{\mu}\hat{k}-\hat{k}\gamma_{\mu}\hat{q}
+2 i m k_{\mu}\}$ 
and  $$\frac{1}{(k^2+\lambda^2)(k^2-2 p_1 k)(k^2- 2 p_2 k)}$$.

        Similarly,   those  in  the  second  curly  bracket   are 
respectively $\{-2m^2\gamma_{\mu}-\hat{k}\gamma_{\mu}\hat{k}
+i m (\gamma_{mu}\hat{k}-\hat{k}\gamma_{\mu}\}$                    
and  $(k^2+\lambda^2)[(p_2-k)^2+m^2][(p_1-k)^2+m^2]$.

        We may write (\ref{eq8})$_a$ as
\renewcommand{\theequation}{A\arabic{equation}}
\setcounter{equation}{3}
\begin{equation}
\big(\left<f\right| S^{(3)} \left|i\right>\big)_a=
\frac{-i e^3}{2} \bar{u}^{(r_2)}(\vec{p}_2)
\int {\mathrm d}^4 k [(I)_a+(II)_a] a_{\mu}(q)u^{(r_1)}(p_1)
\label{eqA4}
\end{equation}
where
\renewcommand{\theequation}{A\arabic{equation}I}
\setcounter{equation}{3}
\begin{eqnarray}
&(I)_a=\frac{2\{2m^2\gamma_{\mu}-\hat{q}\gamma_{\mu}\hat{q}
+\hat{k}\gamma_{\mu}\hat{k}-\hat{k}\gamma_{\mu}\hat{q}
+2 i m k_{\mu}\}}
{(k^2+\lambda^2)[(p_2-k)^2+m^2][(p_1-k)^2+m^2]}
\delta^4(p_1-p_2+q)
\nonumber \\
&=4\int_0^1{\mathrm d} x \int_0^1 {\mathrm d} y
\nonumber \\
&\frac{\{(2-2x-x^2)m^2-\frac{k^2}{2}+(1-x+y)(1-y)q^2\}\gamma_{\mu}
+im q_{\mu}(1+x)(2 y-x) + mx (1-x) \sigma_{\mu\nu}q_{\nu}}
{\{k^2+m^2x^2+q^2y(x-y)+\lambda^2(1-x)\}^3}
\label{eqA4I}
\end{eqnarray}
by usual calculation; here, $\sigma_{\mu \nu}$
\renewcommand{\theequation}{A\arabic{equation}II}
\setcounter{equation}{3}
\begin{eqnarray} 
&(II)_a=\frac{2\{-2 m^2 \gamma_{\mu}-\hat{k}\gamma_{\mu}\hat{k}
+i m (\gamma_{\mu}\hat{k}-\hat{k}\gamma_{mu})\}}
{(k^2+\lambda^2)[(p_2-k)^2+m^2][(p_1-k)^2+m^2]}
\delta^4(p_1-p_2)
\nonumber \\
&=4\int_0^1 {\mathrm d} x \int_0^1 {\mathrm d} y
\frac{\{-(2-2x+x^2)m^2+\frac{k^2}{2}\}\gamma_{\mu}}
{\{k^2+m^2x^2+\lambda^2(1-x)\}^3}
\label{eqA4II}
\end{eqnarray} 
Here, in the case (II),  $x'_1=-x_1$,  
the corresponding $p_1$ should be 
turned  into $-p_1$, (see (\ref{eq1a})), 
Dirac Eq. for $-p(-\vec{p},-E)$ case should be 
$(-i \hat{p}+m)u=0$ in place of
$(i \hat{p} +m) u=0$ for $p(\vec{p},E)$ case.

\renewcommand{\theequation}{A\arabic{equation}}
\setcounter{equation}{5}
\begin{eqnarray}
&&(\ref{eq8})_a=-ie^3\bar{u}^{(r_2)}(\vec{p}_2)
\int {\mathrm d}^4 k \int_0^1 {\mathrm d} x
\int_0^1 {\mathrm d} y 
\nonumber \\
&&\{
\frac{\gamma_{\mu}(1-x+y)(1-y)q^2+mx(1-x)\sigma_{\mu\nu}q_{\nu}}
{(k^2+l^2)^3}
 +\gamma_{\mu}[(2-2 x-x^2)m^2-\frac{k^2}{2}]
\nonumber \\
&&[\frac{1}{(k^2+l^2)^3}-\frac{1}{(k^2+l_0^2)^3}]
-\frac{2 m^2 x^2}{(k^2+l^2_0)^3}\gamma_{\mu}\}
\label{eqA6} 
\end{eqnarray}
where, $l^2=m^2x^2+q^2y(x-y)+\lambda^2(1-x)$, 
$l_0^2=m^2x^2+\lambda^2(1-x)$,   
the Lorentz condition $q_{\mu}a_{\mu}(q)=0$ has been used.

        Here  in  (\ref{eqA6}), there are only the first  terms  in  each 
curly bracket in (\ref{eq8}) , so for the whole (\ref{eq8}), we have
\begin{eqnarray}
&&(\ref{eq8})=-2ie^3\bar{u}^{(r_2)}(\vec{p}_2)
\int {\mathrm d}^4 k \int_0^1 {\mathrm d} x
\int_0^1 {\mathrm d} y 
\nonumber \\
&&\{
\frac{\gamma_{\mu}(1-x+y)(1-y)q^2+mx(1-x)\sigma_{\mu\nu}q_{\nu}}
{(k^2+l^2)^3}
+\gamma_{\mu}[(2-2 x-x^2)m^2-\frac{k^2}{2}]
\nonumber \\
&&[\frac{1}{(k^2+l^2)^3}-\frac{1}{(k^2+l_0^2)^3}]
-\frac{2 m^2 x^2}{(k^2+l^2_0)^3}\gamma_{\mu}\}
\label{eqA7} 
\end{eqnarray}
The last term is just $-\pi^2 i \gamma{\mu}$, 
a first order vertex.  The other terms in 
(\ref{eqA7}),  denoted as (\ref{eq8}'), can be, 
in the case of $q<<m$, expanded  in 
power series in $q$.  We have then, to the order of $q^2$, 
\begin{eqnarray}
(\ref{eq8}')&\cong& \pi^2 e^3 \int_0^1 x {\mathrm d} x
\{\frac{q^2x^2\gamma_{\mu}}{6 l_0^2}
-\frac{q^2m^2x^2}{6 l_0^4}(2-2 x-x^2)\gamma_{\mu}
\nonumber \\
&&
+\frac{q^2}{l_0^2}(1-x+\frac{x^2}{6})\gamma_{\mu}
+\frac{m}{l^2_0}x(1-x)\sigma_{\mu\nu}q_{\nu}\}
\label{eqA8}
\end{eqnarray}
On using the well known integrals
$\int_0^1\frac{{\mathrm d} x}{l_0^2}$,
$\int_0^1\frac{x {\mathrm d} x}{l_0^2}$,
$\int_0^1\frac{x^{n \geq 2}{\mathrm d} x}{l_0^2}$;
$\int_0^1\frac{{\mathrm d} x}{l_0^4}$,
$\int_0^1\frac{x {\mathrm d} x}{l_0^4}$,
$\int_0^1\frac{x^{n \geq 2}{\mathrm d} x}{l_0^4}$ 
and taking $\lambda \to 0$, we have, by ordinary calculations,
\begin{equation}
\Lambda^{(2)}_{\mu f}(p_1,p_2;q)=-\frac{\alpha}{\pi}
\{(\frac{q^2}{3 m^2}\ln \frac{m}{\lambda}-\frac{q^2}{8 m^2})
\gamma_{\mu}-\frac{i}{8 m}(\gamma_{\mu}\hat{q}-\hat{q}\gamma_{\mu})\}                        
\label{eqA9}
\end{equation}
to the order $q^2$, which is just Eq.~(\ref{eq9}).

\end{document}